\documentclass[amsfonts, amssymb, amsmath, reprint, showkeys, nofootinbib, aps, prapplied, superscriptaddress]{revtex4-2}
\usepackage[bottom]{footmisc}
\usepackage{graphicx}
\usepackage{caption}
\usepackage{subcaption}
\usepackage{nicefrac, xfrac}
\usepackage{array}
\usepackage{dcolumn}
\usepackage{bm}

\begin{document}

\preprint{APS/123-QED}

\title{Countermeasure against detector-blinding attack with estimation of secret-key leakage}

\author{Dmitry M. Melkonian}
    \email{Contact author: dmitry.melkonian@gmail.com}
    \affiliation{SFB Laboratory, LLC, 127273 Moscow, Russia}
\author{Daniil S. Bulavkin}
    \email{Contact author: daniel.bulavkin@gmail.com}
    \affiliation{SFB Laboratory, LLC, 127273 Moscow, Russia}
\author{Kirill E. Bugai}
    \affiliation{SFB Laboratory, LLC, 127273 Moscow, Russia}
\author{Kirill A. Balygin}
    \affiliation{Lomonosov Moscow State University, 119991 Moscow, Russia}
\author{Dmitriy A. Dvoretskiy}
    \affiliation{SFB Laboratory, LLC, 127273 Moscow, Russia}
\date{\today}

\begin{abstract}
We present a countermeasure against the detector blinding attack (DBA) utilizing statistical analysis of error and double-click events accumulated during a quantum key distribution session under randomized modulation of single-photon avalanche diode (SPAD) detection probabilities via gate voltage manipulation. Building upon prior work demonstrating the ineffectiveness of this countermeasure against continuous-wave (CW) DBA, we extend the analysis to evaluate its performance against pulsed DBA. Our findings reveal an approximately 13 dB increase in the trigger pulse energies difference between default and reduced gate voltage applied under pulsed DBA conditions compared to CW DBA. This heightened difference enables a re-evaluation of the feasibility of utilizing SPAD detection probability variations as a countermeasure and makes it possible to estimate the fraction of bits compromised by an adversary during pulsed DBA.

\end{abstract}

\maketitle

\section{Introduction}
Quantum key distribution (QKD) leverages fundamental principles of quantum mechanics, notably the Wootters-Surek theorem on the impossibility of cloning a quantum state \cite{wootters1982single}, to provide a theoretical framework for secure communication. This security is predicated on the inherent impossibility of perfectly replicating an unknown quantum state, thus rendering eavesdropping detectable regardless of the computational capabilities of a potential violator, often denoted as Eve. However, practical realizations of QKD systems can be vulnerable to a wide range of attacks that exploit imperfections in the employed hardware (Trojan horse \cite{gisin2006trojan}, Laser damage \cite{bugge2014laser, alferov2022study}, Detector blinding \cite{lydersen2010hacking, sauge2011controlling}, Laser seeding \cite{huang2019laser}, Aftergate \cite{wiechers2011after}, etc \cite{makarov2006effects, qi2005time, lydersen2011controlling, jouguet2013preventing, bogdanov2022influence}). These vulnerabilities compromise the intended security and necessitate introducing countermeasures to detect an adversary, assess and reduce its knowledge of the distributed key \cite{sushchev2024trojan, bugai2024protection, lovic2023quantified}.

This paper is dedicated to improving further the methods for assessing and reducing the influence of a potential eavesdropper on the distributed key. Specifically, we consider the detector blinding attack (DBA) \cite{lydersen2010hacking, lydersen2010thermal, makarov2009controlling, sauge2011controlling, bulavkin2023study, wu2020hacking} on QKD systems utilizing single-photon detectors built upon gated InGaAs/InP avalanche photodiodes (APDs). 

The DBA begins with Eve intercepting and measuring quantum states from Alice. Subsequently, an eavesdropper directs high-power optical radiation at the legitimate receiver's (Bob's) single-photon avalanche diodes (SPADs). The violator exploits the fact that intense illumination can shift these detectors from Geiger mode (single-photon sensitive mode) into a linear operational mode. The radiation can be injected continuously (CW DBA) or via discrete optical pulses applied between the SPAD's gates (pulsed DBA) \cite{wu2020hacking}. Due to this vulnerability, they lose their single-photon sensitivity and can be triggered at a precisely known level of classical optical power. This allows a potential violator to set the trigger pulse energy level such that the in-gate detection events can be triggered only when Eve's chosen measurement bases align with Bob's and full trigger-pulse energy hits Bob's SPAD. This eavesdropping ability crucially jeopardizes widely used QKD systems based on "prepare-and-measure" protocols.

To address this challenge, several solely hardware-based countermeasures were introduced \cite{yuan2011resilience, lydersen2010hacking, chistiakov2019controlling, qian2019robust, wu2020robust, wu2020hacking}. Despite proving to be effective countermeasures against the CW DBA monitoring the current of the SPAD \cite{yuan2011resilience}, the use of the watchdog detector \cite{lydersen2010hacking, chistiakov2019controlling, acheva2023automated} and using variable optical attenuators \cite{qian2019robust, wu2020robust} failed to protect the system from the pulsed DBA. Existing practical countermeasures against pulsed DBA typically necessitate modifications to the receiver's (Bob's) optical setup. For instance, photocurrent monitoring utilizing detector waveform analysis \cite{wu2020hacking} has effectively detected DBAs in both CW and pulsed regimes. However, the practical implementation of detecting weak and fast SPADs' avalanche current changes with an oscilloscope is associated with significant complexity and cost. 

Several studies have also explored the revealing of both types of DBA through the statistical analysis of detector clicks. The study by Shen et al. \cite{shen2025countering} presents an alternative countermeasure employing detector self-testing. This method utilizes an additional optical component, a quasi-single-photon source within Bob's apparatus, and its activation during randomly selected detection gates. The presence of DBA is determined by analyzing detector counts corresponding to these activation periods. A particularly promising approach involves the variation of the SPAD's detection probability \cite{lim2015random}, which can be achieved via supply voltage change \cite{legre2018apparatus}, total voltage level set on the SPAD. This countermeasure offers implementation in higher-frequency systems compared to variable optical attenuators and does not require supplementary optical components. The core principle of the defense is to introduce additional uncertainty for Eve regarding the receiver's system state. This uncertainty aims to force Eve to leave detectable "fingerprints" when attempting to impose clicks. As suggested by the authors, this method requires a comparison between the expected click statistics for each detection probability under normal operating conditions and those observed experimentally to detect Eve's presence. A. Huang et al. \cite{huang2016testing} tested the implementation of this countermeasure by varying the SPADs' bias voltage values or suppressing the detector's gate while conducting CW DBA. They claimed that if an adversary applies a trigger pulse within the gate instead of sending it slightly after it, the countermeasure developed by ID Quantique is vulnerable to such an attack. It was demonstrated that Eve can manipulate the detector click statistic by being able to trigger detectors only at a higher supply voltage. At the same time, the imposition attempts cause no clicks when Eve and Bob's bases mismatch, regardless of the supply voltage level applied.

All the countermeasures mentioned above operate as attack notification systems only, meaning that if the eavesdropper's presence is detected, the whole distributed key should be discarded. The "prepare and measure" QKD protocol offered by S.N. Molotkov \cite{kulik2017decoy} combined with the receiver's measurement system modification \cite{molotkov2021blinding} can theoretically still ensure a successful detection of DBA conducted and an assessment of the fraction of bits imposed by Eve. However, the practical implementation of this countermeasure presents significant challenges and requires the specific QKD protocol to be applied. ID Quantique \cite{bussieres2020blinding, gras2021countermeasure} also designed a multi-pixel superconducting nanowire single-photon detector, which was tested and proved its efficiency for countering DBA. The detector's click statistic analysis also allows for assessing the fraction of imposed bits. Nevertheless, most modern commercially available QKD systems still use common SPADs.

Aiming at finding an easily implementable (no extra Bob's optical setup changes required), yet effective solution, we present another approach based on detector clicks statistic analysis. Our work demonstrates the effectiveness of the countermeasure against the pulsed DBA. Our approach develops an idea of randomizing the receiver's side detection probabilities by adjusting the supply voltage \cite{lim2015random, legre2018apparatus}, specifically through variations in the gate voltage value. We claim that the full distinguishability between trigger pulse energies' ranges for different supply voltages applied can be used in favor of legitimate users. In particular, when Bob varies the gate voltage level, Eve must possess prior knowledge of its value at each detection time slot in order to correctly predict the detector response to the injected trigger pulses under the applied blinding conditions. In the absence of such knowledge, she will inevitably leave detectable “fingerprints,” in particular double-click and error events. For example, to ensure detector clicks when the supply voltage is lower, Eve would need to increase the trigger pulse energy. However, during time slots in which a higher supply voltage is applied, the split energy of such pulses (in the case of a basis mismatch) may still produce a current exceeding the comparator threshold, thereby causing unintended detection events.

To support this claim, we measured the trigger pulse energies corresponding to default and reduced gate voltages using a commercially available SPAD. Our measurements under pulsed DBA reveal a substantial energy difference of more than approximately 13 dB between the lowest trigger pulse energy that always produces a detector click ($E_{always}$) at higher supply voltage and the highest trigger pulse energy insufficient to cause a click ($E_{never}$) at lower supply voltage. Consequently, although Eve may still gain partial control over Bob’s detectors, double-click or error events occur whenever she fails to correctly guess the applied gate voltage level and Bob’s chosen basis. Hence, the observed energy gap enables us to quantify the fraction of clicks imposed under pulsed DBA. By contrast, the original approach \cite{legre2018apparatus} cannot be used either to detect or to quantify an eavesdropper’s influence on the distributed key.

\section{Experimental Setup}

\begin{figure*}[t]
\centering
\fbox{\includegraphics[width=0.9\textwidth]{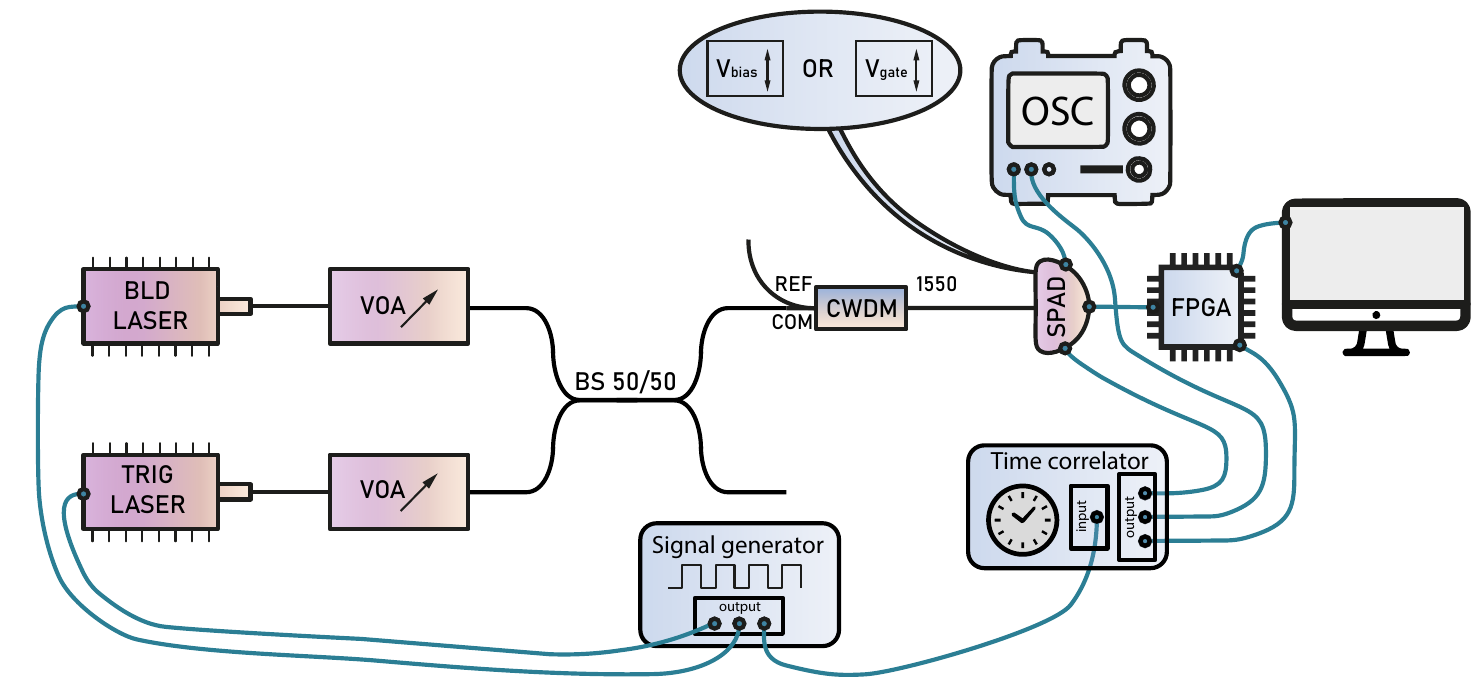}}
\caption{Experimental setup.}
\label{fig:3}
\end{figure*}

The experimental setup is shown in Figure \ref{fig:3}. For our tests, we use a commercially available single-photon detector based on gated InGaAs/InP APD. Custom-designed electronic circuit enables precise changing of both bias and gate voltages.  The blinding and trigger lasers (BLD LASER and TRIG LASER on the scheme, respectively) are semiconductor laser sources with a 1550 nm wavelength. The first one can operate in both continuous and pulsed regimes and performs switching SPAD from Geiger to linear detection mode. The second one served dual purposes: as a reference light source for the measurement of detection probabilities and for triggering the SPAD in the blinded state. The lasers' power is controlled using variable optical attenuators (VOA). Coarse wavelength division multiplexer (CWDM) was implemented to mitigate amplified spontaneous emission (ASE) from the laser sources. The required clock signals, their delays and frequencies are set by the Signal generator independently for both laser sources. A 10~MHz reference clock, derived from the signal generator, was distributed to the SPAD under test, an oscilloscope (OSC), and a field-programmable gate array (FPGA) via a time correlator. The oscilloscope was utilized for real-time monitoring of the SPAD's avalanche signals. The FPGA performed a statistical analysis of click events.

\section{Results} \label{results}

\begin{table*}[htbp]
\caption{Detection probabilities versus supply voltage for the unblinded SPADs.}
\label{tab:detection_probabilities}

\small
\setlength{\tabcolsep}{3pt}
\renewcommand{\arraystretch}{1.25}
\setlength{\extrarowheight}{1.5pt}

\begingroup
\renewcommand{\thefootnote}{\fnsymbol{footnote}}

\begin{ruledtabular}
\begin{tabular*}{\textwidth}{@{\extracolsep{\fill}}c|cccccccc}
\parbox[c]{0.10\textwidth}{\centering Supply\\voltage} &
\parbox[c]{0.10\textwidth}{\centering Default\\settings\footnotemark[1]} &
\parbox[c]{0.09\textwidth}{\centering Reduced Bias:\\$59.65\,\mathrm{V}$} &
\parbox[c]{0.09\textwidth}{\centering Reduced Bias:\\$59.25\,\mathrm{V}$} &
\parbox[c]{0.09\textwidth}{\centering Reduced Gate:\\$3.20\,\mathrm{V}$} &
\parbox[c]{0.09\textwidth}{\centering Reduced Gate:\\$2.95\,\mathrm{V}$} &
\parbox[c]{0.09\textwidth}{\centering Reduced Gate:\\$2.80\,\mathrm{V}$} &
\parbox[c]{0.09\textwidth}{\centering Reduced Gate:\\$2.65\,\mathrm{V}$} &
\parbox[c]{0.09\textwidth}{\centering Reduced Gate:\\$2.50\,\mathrm{V}$} 

\\[10pt]
\hline
\parbox[c]{0.10\textwidth}{\centering $P_{\mathrm{det}},\%$\footnotemark[2]} &
$12.1$ & $6.95$ & $1.95$ & $10.8$ & $8.4$ & $6.9$ & $4.3$ & $1.95$ \\
\end{tabular*}
\end{ruledtabular}

\footnotetext[1]{Bias: $60.5\,\mathrm{V}$; Gate: $3.65\,\mathrm{V}$.}
\footnotetext[2]{Detection probability.}

\endgroup

\renewcommand{\arraystretch}{1}
\setlength{\extrarowheight}{0pt}
\end{table*}

In this section, we compare how reducing either the bias voltage or the gate voltage affects the click probability under both types of DBA. Here, the click probability is defined as the probability that the voltage surge induced by a trigger pulse exceeds the comparator’s predefined threshold. Finally, we contrast these results with the system’s detection probabilities under normal (i.e., unblinded) SPAD operation to highlight the pronounced difference in detector response under DBA. A detailed explanation of the observed behavior under both CW and pulsed DBA, together with a theoretical analysis of the SPAD response under different supply voltage settings, are provided in Section~\ref{attack description}.

First of all, we have carried out multiple series of experiments to measure detection probabilities of SPAD in Geiger mode (no DBA conducted) under various supply voltages. The supply voltage change is achieved by adjusting either the gate or bias voltage values. The results are presented in the Table \ref{tab:detection_probabilities}. As evident from the table, reductions in detection probability achieved by lowering the bias voltage are comparable to those obtained by decreasing the gate voltage when no DBA is applied.

\begin{figure*}[htbp]
\centering
\includegraphics[width=0.90\textwidth]{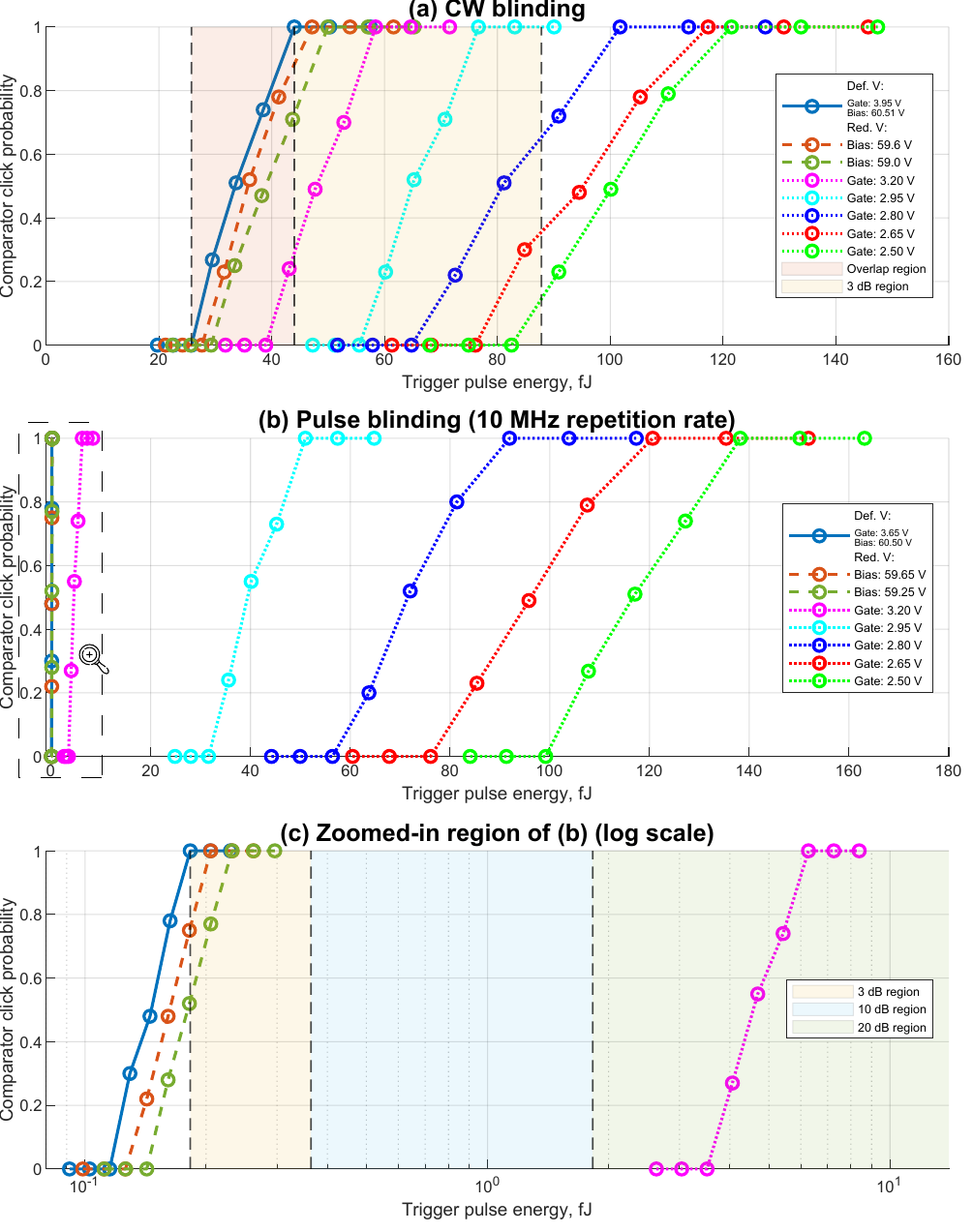}
\caption{Comparator click probability versus trigger pulse energy under CW (a) and 10 MHz pulsed blinding (b, c). The color coding of the curves is identical in panels (a)–(c). Shaded areas indicate the corresponding energy regions relative to $E_{always}^{Def. \, V}$. Notes "Def. V" and "Red. V" state for default and reduced supply voltage settings respectively according to the Table \ref{tab:detection_probabilities}}
\label{fig:4}
\end{figure*}

Subsequently, we have measured the comparator's click probability across a range of trigger pulse energies under both CW (Figure \ref{fig:4}a) and pulsed (Figures \ref{fig:4}b and \ref{fig:4}c) DBA conditions for identical sets of gate and bias voltages. The pulsed DBA case measurements were also performed across several blinding pulse repetition rates, ranging from 2 to 10 MHz, to test the detection system's response to different mean blinding powers.

\begin{figure*}[htbp]
\centering
\includegraphics[width=0.9\textwidth]{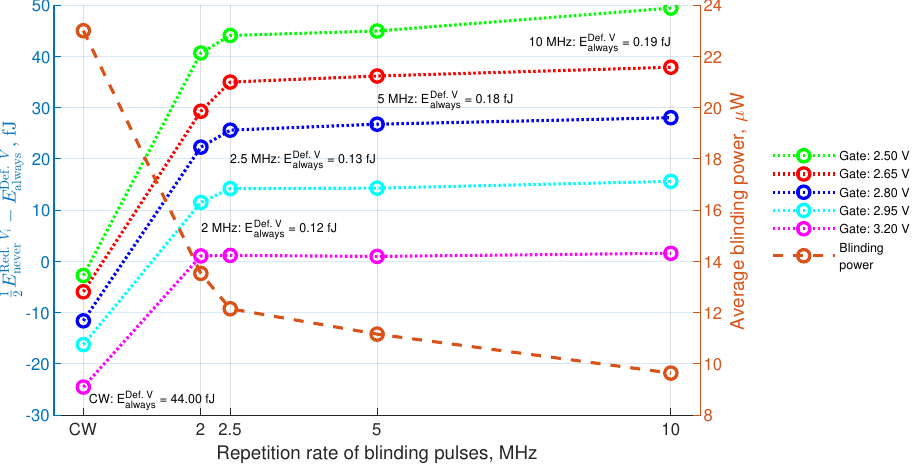}
\caption{Testing the condition (\ref{eq:refname1}) for leaving "fingerprints" under pulsed (for several blinding pulse repetition rates) and CW DBA. We compare 
the default-gate-applied (3.65V) case and several lower ones. The average blinding power versus the repetition rate curve indicates correlations between the blinding power and the trigger-pulses energy gap.}
\label{fig:5}
\end{figure*}

We first consider the CW DBA case (Figure \ref{fig:4}a). When the gate voltage is changed, the energy curves corresponding to [$E_{never}$, $E_{always}$] intervals for the reduced and default gate voltage settings either exhibit minimal overlap or are separated insufficiently (all $E_{never}$ values for lowered supply voltage are within the 3 dB region in \ref{fig:4}a). As mentioned above, this allows Eve to impose bits without leaving detectable “fingerprints” in the detection statistics \cite{huang2016testing}. In turn, varying the bias voltage is less effective in separating the energy curves associated with the default and lowered bias settings. Thus, similarly to the gate-voltage manipulation case, Eve can choose trigger pulse energies that produce a detector click only when Bob’s basis is correctly anticipated. If the bases do not match, the divided trigger pulse energy remains insufficient to generate a detection event - even if Eve incorrectly guesses the APD supply voltage level.

Variation of the bias voltage under pulsed DBA (orange and green curves in Figure \ref{fig:4}c) likewise does not produce a significant separation between the energy ranges [$E_{never}$, $E_{always}$] corresponding to the default and reduced supply voltages. These intervals overlap similarly to the CW DBA case, resulting in the same vulnerability. However, when the supply voltage is lowered via gate voltage level alteration under the pulsed DBA, the trigger pulse energy should be increased by more than 13 dB (Figures \ref{fig:4}b and \ref{fig:4}c) to impose a click for the detector in a lower supply voltage mode. This highlights the difference in detector response to supply-voltage adjustment methods in Geiger mode (see Table~\ref{tab:detection_probabilities}) compared with pulsed blinding.

This increased separation of the energy intervals [$E_{never}$, $E_{always}$] renders the presence of an eavesdropper quantitatively assessable. In most QKD protocols, when Eve’s chosen basis does not match Bob’s, the energy of the pulse reaching each SPAD is half of the trigger pulse energy injected by Eve. For an adversary’s intervention to become detectable, these halved pulse energies must exceed $E_{always}$ corresponding to the default supply voltage ($E_{always}^{Def. V}$). Accordingly, the condition for leaving such detectable “fingerprints” can be expressed as
\begin{equation}
E_{always}^{Def. \, V} \leq \frac{1}{2}E_{never}^{Red. \, V}.
\label{eq:refname1}
\end{equation}
The substantial difference between $E_{never}^{Red. \, V}$ ($E_{never}$ corresponding to the reduced supply voltage) and $E_{always}^{Def. \, V}$ inevitably results in simultaneous clicks of both detectors if Eve attempts to enforce clicks while incorrectly guessing both Bob’s measurement basis and the selected reduced-voltage (gate) setting.

Figure \ref{fig:5} demonstrates our test of the satisfaction (positive values for expression $\frac{1}{2}E_{never}^{Red. \, V} - E_{always}^{Def. \, V}$) of the condition (\ref{eq:refname1}) depending on the applied type of DBA and on blinding pulses repetition rate (under pulsed DBA). Values of $E_{always}^{Def. \, V}$ are shown to highlight the significant separation between the energy intervals ensuring that (\ref{eq:refname1}) can be easily met with a large margin under pulsed DBA.

It is also worth noting that the gap between $E_{always}^{Def. \, V}$ and $E_{never}^{Red. \, V}$ correlates with the average blinding power (and resulting photocurrent) we applied during our experiments, as shown in Figure \ref{fig:5}. This gap widens as the blinding pulse frequency increases, which corresponds to a decrease in the average blinding power level. However, CW DBA almost negates the effect due to a significant photocurrent increase through APD, resulting from the overlap of the applied blinding radiation with the SPAD's gate.

Consequently, the system with the countermeasure applied still necessitates support of some basic detector current control \cite{yuan2011resilience}. The alarm threshold for the current monitoring system needs careful calibration to ensure that inevitable fluctuations in the APD's circuit temperature and supply voltage do not violate (\ref{eq:refname1}). Our experiment demonstrates  the 10+ dB between $\frac{1}{2}E_{never}^{Red. \, V}$ and $E_{always}^{Def. \, V}$ even for slight reduction of the gate voltage (and thus, non-zero detection probability in Geiger mode), confirming the countermeasure's potential to facilitate secret key generation even under the pulsed DBA. However, it's important to note that reducing the supply voltage, while beneficial against DBAs, concurrently decreases the detection probability during normal operation, thereby lowering the secret key rate (SKR). Therefore, maintaining the supply voltage as close as possible to its normal operating level is preferable. Nevertheless, inequality (\ref{eq:refname1}) imposes constraints on the permissible range of supply voltage settings.

\section{Theoretical analysis of SPAD's response under DBA} \label{attack description}

This section elucidates the distinctions between changing SPADs' operation modes via bias and gate voltage manipulation, and the differences in APD responses on trigger pulses injected by Eve under CW and pulsed DBAs.

\begin{figure}[htbp]
\centering
\includegraphics[width=0.45\textwidth]{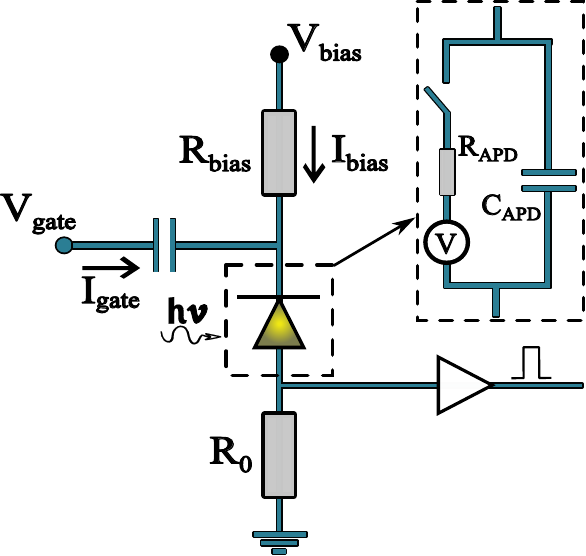}
\caption{Considered passive-quenching SPAD's circuit and the APD's equivalent scheme. $C_{APD}$ and $R_{APD}$ denote the APD's inner capacity and resistance. The APD's equivalent circuit closes under blinding light. The detailed description is presented in \cite{cova1996avalanche}.}
\label{fig:1}
\end{figure}

When Eve injects light in order to blind Bob's SPADs, it initially triggers an avalanche, increasing the current. Although the current level rapidly decreases after the initial avalanche, Eve can maintain a substantial current by persistently applying classical radiation to SPADs. The blinding power is selected not to induce a photocurrent increase (and therefore voltage surge on $R_{0}$ (Figure \ref{fig:1})) significant enough to surpass the comparator's threshold and cause a detection event.

The current generated by the blinding light passes through the APD and the avalanche quenching resistor $R_{bias}$ (Figure \ref{fig:1}). Following Ohm's Law, the resulting voltage increase on $R_{bias}$ substantially reduces the APD's supply voltage. This shifts the APD into an operational mode characterized by a significantly lower avalanche multiplication factor (M), whose dependence on the supply voltage $U_{sup}$ and breakdown voltage $U_{br}$ can be described by the empirical Miller formula for InAsGa/InP APDs \cite{ma2002multiplication, miller1955avalanche}:
\begin{equation}
M = \frac{1}{1 - (\frac{U_{sup}}{U_{br}})^{n}},
\label{eq:mfactor}
\end{equation}
and therefore is directly determined by the effective voltage across the APD.

Thus, the supplied voltage under DBA (and, therefore, the M-factor) can be reduced in three ways: (i) by reducing the bias voltage; (ii) by reducing the gate voltage; or (iii) by increasing the blinding power. 

Now let us compare the impact of decreasing the bias voltage versus decreasing the gate voltage on the change in supply voltage, $\triangle$V, under the same blinding level.  When the bias voltage $V_{bias}$ is decreased by $\triangle$V, the current through the $R_{bias}$ -- $R_{APD}$ circuit drops by:
\begin{equation}
\triangle I_{bias} \approx \frac{\triangle V}{R_{bias} + R_{APD}}.
\label{eq:refname2}
\end{equation}

It is reasonable to assume that $R_{bias}$ $\gg$ $R_{APD}$ $\gg$ $R_{0}$ (typically $R_{bias}\approx$10 k$\Omega$, $R_{APD}\approx$400 $\Omega$, $R_{0}\approx$50 $\Omega$) due to two primary reasons. First, the voltage surge on $R_{bias}$ must be significantly larger than that across $R_{APD}$ to ensure effective avalanche quenching. Second, the resistance $R_{0}$ is chosen to be negligibly small, as its sole purpose is to provide photocurrent measurements. Therefore, if $V_{bias}$ decreases by $\triangle $V the APD's supply voltage decreases by:
\begin{equation}
\triangle V_{biasAPD} \approx \frac{\triangle V R_{APD}}{R_{bias} + R_{APD}}.
\label{eq:refname3}
\end{equation}

At the same time when the gate voltage $V_{gate}$ is decreased by $\triangle$V, the APD's supply voltage is also reduced by $\triangle$V.

\begin{figure}[htbp] 
\centering
\includegraphics[width=0.45\textwidth]{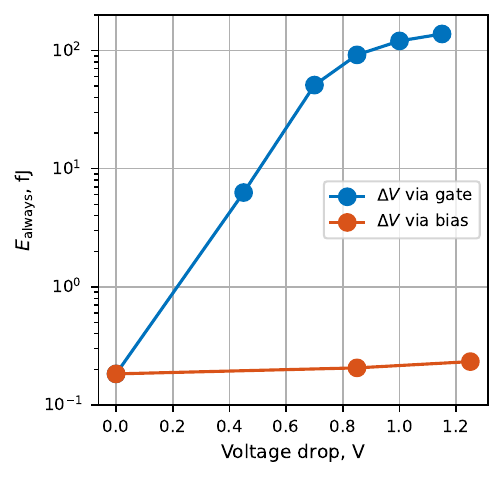}
\caption{Dependence of the \(E_{always}\) on the voltage drop introduced through gate and bias control. The voltage reduction modifies the effective avalanche multiplication factor, thereby altering the detector response under pulsed blinding.}
\label{fig:8}
\end{figure}

Taking into account that $R_{bias}$ $\gg$ $R_{APD}$ we get:
\begin{equation}
\triangle V \gg \triangle V_{biasAPD}.
\label{eq:refname4}
\end{equation}
The expressions (\ref{eq:refname3}) and (\ref{eq:refname4}) yield several key conclusions. First, the APD's supply voltage under blinding in the case of decreasing bias voltage remains almost the same as the default one. Therefore, the shifted $E_{always}$ and $E_{never}$ energy levels are also close to the ones before the bias level change. Second, varying the gate voltage level appears to be a significantly more effective method for increasing the divergence between supply voltage levels on the APD and thus the difference between $E_{always}^{Red. \ V}$ and $E_{never}^{Def. \ V}$. Experimental confirmation of this effect is provided in Figure \ref{fig:8}. Here, we focus specifically on the 100$\%$ click probability level because it can be determined with significantly higher precision than intermediate probability values.

\begin{figure}[htbp] 
\centering
\includegraphics[width=0.445\textwidth]{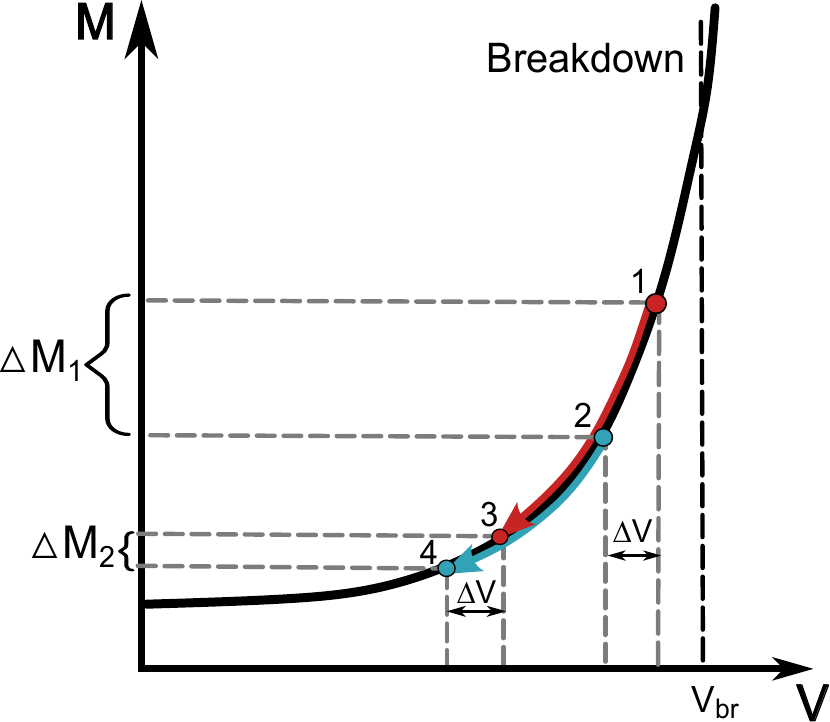}
\caption{Avalanche Multiplication Factor (M) vs. APD Supply Voltage (V).  Red and light blue arrows, along with dashed lines, illustrate the change in M-factor difference between default and lowered supply voltages, respectively, as blinding light power increases.}
\label{fig:2}
\end{figure}

Moreover, the initial rapid decrease in the M-factor implies that even a small change (0.5 - 1 V) in the APD's supply voltage substantially increases the energy difference of trigger pulses for default and reduced gate levels under low power (pulsed) DBA (see $\Delta M_{1}$ between points 1 and 2 in Figure \ref{fig:2}).

Importantly, as the multiplication factor decreases, its sensitivity to further reductions in the supplied voltage progressively weakens due to the nonlinear dependence described by (\ref{eq:mfactor}). In other words, once the detector operates at a sufficiently reduced M (e.g., point 4 and below in Figure~\ref{fig:2}), additional decreases in the supply voltage - regardless of the method - lead to progressively smaller (in relative units) changes in the avalanche amplitude. Consequently, the growth of $E_{always}$ and $E_{never}$ becomes increasingly limited.

Finally, we analyze the reduction in the difference between the trigger pulse energy levels as the applied optical radiation increases (see the shrinkage of the [$E_{always}^{Def. \, V}$, $E_{never}^{Red.\, V}$] ranges with increasing average blinding power in Figure ~\ref{fig:5}). This reduction arises from the diminishing variation of the APD multiplication factor 
M between the default and reduced supply-voltage regimes -- whether the reduction is achieved via gate or bias adjustment -- under elevated blinding power (see points 3 and 4 in Figure \ref{fig:2}). Specifically, comparing $\Delta M_{2}$ (corresponding to points 3 and 4 at higher blinding power) with $\Delta M_{1}$ (at lower blinding power) clearly shows that the gap in the multiplication factor decreases as the blinding power increases. The CW DBA case particularly illustrates this effect. In this scenario, the APD current increases not only due to elevated optical power but also because the blinding radiation hits SPADs during the gate, causing an additional current surge in the circuit.

\section{Estimation of the secure bits fraction} \label{fraction estimation}

In this section, we demonstrate that the proposed countermeasure enables quantification of information leakage to a potential eavesdropper in a 2-SPAD-based QKD system and compare the resulting secret key rate~(SKR) with that of the standard decoy state BB84 protocol. A typical Bob configuration for such a system is described in \cite{zhao2006experimental}. Detailed analysis for the 1-SPAD-based implementation is provided in Appendix~\ref{B}, with an example discussed in \cite{kravtsov2018relativistic}.

As discussed above, varying the gate voltage also enables estimation of the number of imposed clicks. We assume that Bob randomly selects either the default or reduced gate voltage for both detectors simultaneously, with probabilities $\alpha$ and $1-\alpha$, corresponding to Geiger-mode detection efficiencies $\eta_1$ and $\eta_2$, respectively. When performing a DBA against such a system, Eve must guess both Bob’s measurement basis and the applied gate level. She therefore sends trigger pulses with $E_{always}$ corresponding to her assumed voltage setting. Let $P_{high}$ and $P_{low}$ denote the probabilities of sending pulses matched to the reduced (higher-energy trigger) and default (lower-energy trigger) gate levels, respectively. To set the upper bound of influenced bits, we assume that Eve conducts a fake-state attack on every bit intercepted from Alice. Thus, let's denote $P_{high} = \beta$ and $P_{low} = 1- \beta$.

Evaluating the number of clicks imposed by an eavesdropper on 2-SPAD-based QKD systems involves several steps. First, Alice and Bob perform the basis reconciliation procedure according to the selected QKD protocol.  Second, Bob sifts the data, discarding all bits where his measurement basis did not match Alice's. Then, Bob calculates the remaining number of double-clicks $N_{double}^{exp}$ experimentally observed during the QKD session (or error-clicks $N_{error}^{exp}$ for 1-SPAD-based QKD systems (see Appendix \ref{B})). All such events are regarded as failed imposition attempts. Finally, the calculation of the fraction of successfully imposed clicks is based on the probability of a double-click event under DBA and the number of double-clicks identified.

Given a 0.5 probability of basis mismatch between Alice and Eve, approximately half of Eve's imposition attempts are discarded during key sifting. Thus, the experimentally determined fraction of double clicks after key sifting ($\nicefrac{N_{double}^{exp}}{N_{sent}}$) represents only half of the "fingerprints" left by Eve. Therefore, according to the experimental results obtained in the previous section and assuming the detectors are identical, the double clicks' imposition probability to the total number of fake-states sent by Eve ($N_{sent}$) can be assessed as:
\begin{equation}
P_{double}^{bld} = 2\frac{N_{double}^{exp}}{N_{sent}} = \frac{\alpha \beta}{2}.
\label{eq:refname5}   
\end{equation}

Also, the probability of successful click imposition ($P_{2}^{suc}$) can be represented by the following expression:
\begin{align}
P_{2}^{suc} =\frac{N_{success}}{N_{sent}} =\frac{\alpha + \beta - \alpha \beta}{2}.
\label{eq:refname6}
\end{align} 

Here we denote the assessed number of successfully imposed clicks as $N_{success}$. The detailed derivation of (\ref{eq:refname5}) and (\ref{eq:refname6}) is provided in Appendix \ref{A}.

The assessed number of fake-states sent by Eve and successfully detected by Bob ($N_{clicked}^{Eve}$) corresponding to the observed number of double clicks is calculated from (\ref{eq:refname5}) and (\ref{eq:refname6}) as follows:

\begin{align}
N_{clicked}^{Eve} = N_{success} + 2N_{double}^{exp} = \frac{1}{2}(\alpha N_{sent} + \frac{4N_{double}^{exp}}{\alpha}).
\label{eq:refname7}
\end{align}

Finally, we calculate the fraction of sifted secure bits after the basis reconciliation procedure ($S_{key}^{2SPAD}$). This is based on the experimentally observed number of clicks obtained by Bob, $N_{clicked}^{exp}$, and is given by:

\begin{equation}
   \begin{aligned}
    S_{key}^{2SPAD} &= \frac{N_{clicked}^{exp} - N_{clicked}^{Eve}}{N_{clicked}^{exp}} =\\
    &= 1 - \frac{1}{2} \biggl( \alpha \frac{ N_{sent}}{N_{clicked}^{exp}} + \frac{4N_{double}^{exp}}{\alpha N_{clicked}^{exp}}\biggr).
    \label{eq:refname8}
\end{aligned} 
\end{equation}

Note that $N_{sent}$ corresponds to the maximum number of bits Eve can send to Bob. Therefore, $N_{sent} = N_{Alice}Q^{Eve}$, where $N_{Alice}$ is the total number of pulses sent by Alice to the quantum channel, $Q^{Eve}$ - Eve's total gain for \{$\mu_0$, $\mu_{1}$, $\mu_{2}$ \} decoy-state protocol \cite{ma2005practical}:

\begin{equation}
Q^{Eve} = \sum \limits _{i=0}^{2} {n_{i}(1 - e^{-\mu_{i}})},
\label{eq:refname9}
\end{equation}
where $n_{i}$ is the probability of Alice preparing each type of state (decoy or legitimate). For a conservative security analysis, we assume Eve can replace the quantum channel with an ideal one exhibiting no transmission losses and achieve perfect detection of Alice's signals ($\eta_{Eve} = 1$), thus increasing the number of states sent to Bob.

The double-click events caused under DBA are intrinsically difficult to distinguish from the ones during an unattacked QKD session. Consequently, the proposed countermeasure must be applied continuously. Thus, it is essential to ensure that this countermeasure does not substantially decrease the SKR when no DBA is present or when Eve successfully mimics the normal single and double-click rates of Bob's SPADs.

We proceed to evaluate the optimal value of $\alpha$ for efficient key generation while minimizing Eve's knowledge of the distributed key under the DBA. The observed values during the QKD session, $N_{clicked}^{exp}$ and $N_{double}^{exp}$, are considered as outputs from a "black box" because the clicks caused by Eve's trigger pulses are indistinguishable from those produced by legitimate states. The maximum of (\ref{eq:refname8}) is achieved for:
\begin{align}
\alpha_{opt} = 2\sqrt{\frac{N_{double}^{exp}}{N_{Alice}Q^{Eve}}}
\label{eq:refname10}
\end{align}
and reaches:
\begin{align}
    S_{key}^{2SPAD} = 1 - 2\frac{\sqrt{N_{double}^{exp}N_{Alice}Q^{Eve}}}{N_{clicked}^{exp}}.
    \label{eq:refname11}
\end{align}

For further numerical simulation of $S_{key}^{2SPAD}$ and SKR we consider: $\{0.6, 0.2, 0\}$ decoy-state BB84 protocol, with decoy-states preparation probabilities $n_{i} =\{0.5, 0.25, 0.25\}$ respectively, the detection probability for higher gate $\eta_1 = 0.121$, dark count rate $Y_{0} = 3*10^{-6}$, transmission losses $\eta_{ch} = 10^{-0.02L}$ (L - channel length). Moreover, we assume that Eve can successfully fake the normal single and double-click rates by applying trigger pulses of different $E_{always}$ energies, as well as by letting several pulses pass from Alice to Bob.

\begin{figure*}[ht]
  \begin{subfigure}[b]{0.48\textwidth}
    \includegraphics[width=\textwidth]{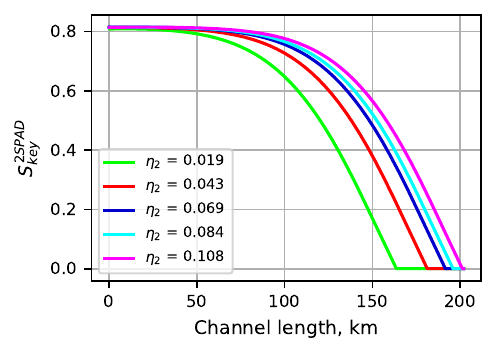}
  \end{subfigure}
  \begin{subfigure}[b]{0.485\textwidth}
    \includegraphics[width=\textwidth]{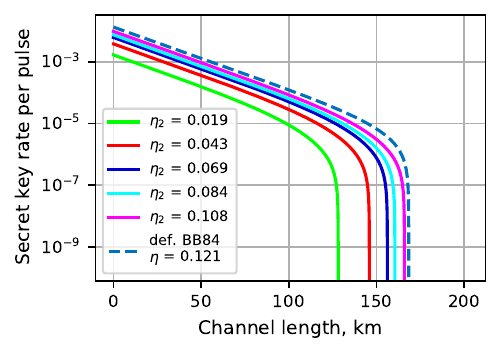}
  \end{subfigure}
  \caption{Secure bit fraction (left figure) and SKR per pulse (right figure) versus QKD channel length for a 2-SPAD QKD system under DBA assuming Eve simulates total and double click gains. Solid curves correspond to the proposed countermeasure for different reduced detection probabilities ($\eta_{2}$). The dotted line shows the BB84 SKR  with the detector gate voltage kept unchanged ($\eta = \eta_{1}$)}
\label{fig:6}
\end{figure*}

Therefore, we anticipate that $N_{clicked}^{exp}$ coincides with the number of clicks Bob could normally obtain without a DBA:
\begin{align}
    N_{clicked}^{exp} = Q^{pass} N_{Alice}.
    \label{eq:refname14}
\end{align}
$Q^{pass}$ - Bob's gain for 2 detectors under normal conditions when legitimate users' bases match:
\begin{align}
    Q^{pass} = \frac{1}{2} \biggl( \alpha \sum \limits _{i=0}^{2} n_{i} Q^{\mu_{i}}_{\eta_{1}} + (1-\alpha) \sum \limits _{i=0}^{2} n_{i} Q^{\mu_{i}}_{\eta_{2}}\biggr),
    \label{eq:refname15}
\end{align}
where $Q^{\mu_{i}}_{\eta_{j}} = 1- (1 - Y_{0})e^{-\mu_{i}\eta_{ch}\eta_{j}}$ - Bob's gain under normal operation conditions for the given $\mu_{i}$ and $\eta_{j}$.

When no DBA is conducted, the observed double-click gain $Q^{double}$ after bases reconciliation procedure is associated with dark counts and optical system's misalignment~$1-T = 0.01$:
\begin{equation}
    \begin{aligned}
        Q^{double} = &\frac{1}{2} \biggl( \alpha \sum \limits _{i=0}^{2} n_{i} Q^{\mu_{i}(1-T)}_{\eta_{1}} Q^{\mu_{i}T}_{\eta_{1}} + \\
        &(1-\alpha)\sum \limits _{i=0}^{2} n_{i} Q^{\mu_{i}(1-T)}_{\eta_{2}}Q^{\mu_{i}T}_{\eta_{2}}\biggr).
    \label{eq:refname16}
    \end{aligned}    
\end{equation}

Thus, considering that $N_{double}^{exp} = Q^{double} N_{Alice}$, the simulation of the assessed fraction of unaffected bits for various $\eta_2$ is presented in the left panel in Figure (\ref{fig:6}).

Accordingly, the lower bound on the key rate, $R^{L}_{DBA}$, can be estimated by Alice and Bob using the modified GLLP formula for the decoy-state method \cite{ma2005security, ma2005practical} under DBA conditions:
\begin{equation}
    \begin{aligned}
        R^{L}_{DBA} = &\frac{1}{2} S_{key}^{2SPAD}\biggl( \mu_{0} e^{-\mu_{0}} Y_{1}^{L}[1-H_{2}(e_{1}^{U})] \\
        &-Q_{\eta_{1},\eta_{2}}^{\mu_{0}} f_{EC} H_2(E_{\eta_{1},\eta_{2}}^{\mu_{0}}) \biggr).
        \label{eq:refname17}
    \end{aligned}
\end{equation}
Here $H_{2}(x)$ is Shannon entropy, $Y_{1}^{L}$ and $e_{1}^{U}$ denote the yield and error rate of the single-photon component, respectively, while $Q_{\eta_{1},\eta_{2}}^{\mu_{0}}$ and $E_{\eta_{1},\eta_{2}}^{\mu_{0}}$ represent the gain and the error-rate of the signal states. These quantities are defined as follows:

\begin{equation}
    \begin{aligned}
        Y_{1}^{L} = &\frac{\mu_{0}}{\mu_{0}\mu_{1}- \mu_{1}^{2}} \biggl( Q_{\eta_{1},\eta_{2}}^{\mu_{1}} e^{\mu_{1}} - Q_{\eta_{1}^,\eta_{2}}^{\mu_{0}} e^{\mu_{0}} \frac{\mu_{1}^{2}}{\mu_{0}^{2}} \\
        &- \frac{\mu_{0}^{2} - \mu_{1}^{2}}{\mu_{0}^{2}}Y_{0}\biggr),\\
        &e_{1}^{U} = \frac{E_{\eta_{1},\eta_{2}}^{\mu_{1}} Q_{\eta_{1},\eta_{2}}^{\mu_{1}} e^{\mu_{1}} - \frac{1}{2}Y_{0} }{Y_{1}^{L} \mu_{1}},\\
        &E_{\eta_{1},\eta_{2}}^{\mu_{i}} = \frac{(1-T)Q_{\eta_{1},\eta_{2}}^{\mu_{i}} + \frac{1}{2}Y_{0}}{Q_{\eta_{1},\eta_{2}}^{\mu_{i}}},\\
        &Q_{\eta_{1},\eta_{2}}^{\mu_{i}} = \alpha Q^{\mu_{i}}_{\eta_{1}} +(1-\alpha)Q^{\mu_{i}}_{\eta_{2}}.
        \label{eq:refname18}   
    \end{aligned}
\end{equation}

Also, in our simulation we put $f_{EC} =1.2$. The simulation of the guaranteed SKR for the legitimate users, together with its comparison to the standard BB84 protocol without the proposed countermeasure (i.e., with the supply voltage kept unchanged by Bob), is presented in the right panel of Figure \ref{fig:6}.

Thus, we have shown that even if a potential eavesdropper exploits the maximum double-clicks' gain achievable under normal conditions to maximize click imposition attempts during lowered gates, Bob can select an $\alpha$ value that preserves over half of the default SKR after privacy amplification for typical QKD channel lengths (100-120 km). Moreover, this approach preserves secret key generation even under detected pulsed DBAs (when the normal double-click rate for the default gate is exceeded), as it enables quantification of the attacked-bit fraction and its subsequent removal during privacy amplification.

We also emphasize that the $\beta$ parameter, which determines the fraction of bits exposed to a DBA, should be calculated from (\ref{eq:refname5}) while considering the finite length of the distributed key. However, it is easy to see, that $\beta$-value asymptotically approaches a normal distribution with a mean of $\sim$ $\nicefrac{N_{double}^{exp}}{N_{sent}}$, because the binomial distribution of $N_{double}^{exp}$ turns into a normal distribution for large numbers. Therefore, the standard deviation of $\beta$ scales as $\sigma_{\beta} \sim \nicefrac{1}{\sqrt{N_{sent}}}$. Consequently, statistical fluctuations have only a negligible impact on the estimated fraction of unattacked bits and SKR.

\section{Soft-attack scenario} \label{soft-attack}

To further emphasize that the attack scenario considered in the previous section represents the most conservative assessment of the defense against Eve's optimal strategy, we now consider a soft-attack scenario. In Section \ref{results}, we argued that Eve's attempts to induce detector clicks in the reduced gates would inevitably generate double-click events in the default gates, and in Section \ref{fraction estimation} we quantified the secure bit fraction while accounting for the number of double-clicks. Here, we instead consider the case in which Eve seeks to avoid producing additional double-click events by applying trigger pulses with energies lower than $2E_{always}^{Def. , V}$. We then estimate the corresponding number of imposed bits and show that Eve's impact on the secure-bit-fraction reduction does not exceed the one obtained in the previous section.

Thus, in the case of a soft attack, Eve operates only within the energy range $2E_{always}^{Def. , V} > E_{Eve} \ge E_{always}^{Def. , V}$ ($P_{high}= \beta = 0$, $P_{low}= 1- \beta = 1$). 
In this regime it is reasonable to partition this range into two subranges:
\begin{enumerate}
    \item $2E_{always}^{Def. , V} > E_{Eve} \ge 2E^{Def. , V}_{never}$;
    \item $2E^{Def. , V}_{never} > E_{Eve} \ge E^{Def. , V}_{always}$.
\end{enumerate}

Considering the first subrange, when Eve's and Bob's bases are 
mismatched, the divided pulse energy can still cause a detector 
click with probability $\gamma$. Thus, $P_{double}^{bld}$ and 
$P_{2}^{suc}$ can be expressed, analogously to the previous 
section and Appendix \ref{A}, as:
\begin{equation}
P_{double,soft}^{bld} = 2\frac{N_{double}^{exp}}{N_{sent}} 
= \frac{\alpha \gamma^2}{2}.
\label{eq:refname19}   
\end{equation}
\begin{equation}
    P_{2,soft}^{suc} = \frac{N_{success}^{soft}}{N_{sent}}
    = \frac{1}{2}\alpha\left(1 + \gamma - \gamma^2\right).
    \label{eq:refname20}
\end{equation}
Obtaining $\gamma$ from (\ref{eq:refname19}) and putting in (\ref{eq:refname20}), the assessed number of fake-states sent by Eve and successfully detected by Bob under soft attack ($N_{clicked, soft}^{Eve}$) corresponding to the observed number of double clicks is calculated as follows:
\begin{equation}
    \begin{aligned}
        N_{clicked, soft}^{Eve} = N_{success}^{soft} + 2N_{double}^{exp} = \\
         =\frac{\alpha N_{sent}}{2}\biggl(1 + 2\sqrt{\frac{N_{double}^{exp}}   {\alpha N_{sent}}}\biggr).
        \label{eq:refname21}
    \end{aligned}
\end{equation}
The secure bit fraction has the same form as (\ref{eq:refname8}), with the substitution $N_{clicked}^{Eve} \rightarrow N_{clicked,soft}^{Eve}$. Thus, to regard our estimate for the number of clicks imposed by Eve as the most conservative one, we compare $N_{clicked}^{Eve}$ (\ref{eq:refname8}) with $N_{clicked,soft}^{Eve}$ (\ref{eq:refname21}). The assessment obtained in Section \ref{fraction estimation} exceeds the one under the soft attack, i.e. $N_{clicked}^{Eve} \geq N_{clicked,soft}^{Eve}$, whenever:
\begin{equation}
    \alpha \leq \alpha_{bound}, \qquad 
    \alpha_{bound} \equiv \left(\frac{4N_{double}^{exp}}{N_{sent}}\right)^{1/3}.
    \label{eq:refname22}
\end{equation}
The optimal value (\ref{eq:refname10}) can be written in the same notation as $\alpha_{opt} = \left(4N_{double}^{exp}/N_{sent}\right)^{1/2}$. Since Eve is forced to keep the double-click gain close to its value under normal operating conditions, the ratio $N_{double}^{exp}/N_{sent} \ll 1$, so that $\alpha_{opt} < \alpha_{bound}$. The estimate of Eve's intervention given in the previous section therefore remains the more conservative one not only for the optimally chosen $\alpha$, but also within a sufficiently wide range around it, since their ratio:
\begin{equation}
    \frac{\alpha_{bound}}{\alpha_{opt}} 
    = \left(\frac{N_{sent}}{4N_{double}^{exp}}\right)^{1/6} \gg 1,
    \label{eq:refname23}
\end{equation}
confirming that $\alpha_{opt}$ lies well inside the admissible interval.

We now turn to the second subrange. When Eve's and Bob's bases are mismatched, the trigger pulse is split equally between the two detectors, and the energy received by each one is insufficient to generate a click, so that $\gamma = 0$. Substituting this value into (\ref{eq:refname20}) and comparing with (\ref{eq:refname6}) shows that the fraction of Eve-imposed bits estimated in Section \ref{fraction estimation} is always greater than or equal to that obtained for the soft-attack scenario, regardless of $\alpha$.

Finally, note that in the full soft-attack scenario, Eve intercepts the entire bit sequence sent by Alice while generating no clicks during the reduced-gate intervals. The resulting decrease in the gain associated with the lower supply voltage may therefore reveal her presence. Eve may instead perform a partial attack, leaving a fraction of Alice’s pulses undisturbed. This further reduces both the number of Eve-induced clicks and the corresponding fraction of Eve-imposed bits in the final key. Since the full soft-attack scenario is already covered by our estimate, any partial attack is necessarily encompassed by the estimates derived in Section \ref{fraction estimation}.

\section{Discussion}
Our measurements were performed by setting a fixed supply voltage level, i.e., by adjusting and then keeping constant either the bias or the gate voltage during each experimental run. Although a practical implementation of the considered approach requires random voltage switching, the use of static voltage settings in our experiments does not compromise the validity of the conclusions. Under the proposed gating strategy, critical changes in detector response in Geiger mode are unlikely to occur in a manner detrimental to the countermeasure as well. To the best of our knowledge, possible deviations in detector behavior are primarily associated with dark count rate and afterpulsing. First, numerous established techniques exist for suppressing afterpulsing in InGaAs/InP SPADs, including reduction of the gate width, active quenching, optimized dead-time, and related approaches \cite{restelli2012time, restelli2013single, losev2022dead}. Second, such effects are generally more pronounced at higher supply voltages and correspondingly higher detection probabilities. In contrast, the proposed countermeasure operates by reducing these parameters, which inherently decreases the probability of "memory effects".

Besides, in linear mode under blinding, the detector's response is governed by the instantaneous applied voltage. While memory effects may still be present due to operation near breakdown \cite{jain2017inp}, the multiplication factor M in this regime is significantly lower than in Geiger mode. Consequently, the avalanche charge -- being the primary driver of afterpulsing -- is substantially reduced. This implies that memory effects in linear mode are further suppressed, including by the mechanisms described above \cite{williams2013multi}. Additionally, imposed avalanche events (and consequently, detector clicks) activate the detector's dead time, which further mitigates potential response deviations by preventing immediate re-triggering.

Moreover, we would like to emphasize that our countermeasure relies on estimating both the fraction of bits affected by blinding and the fraction of bits that could be successfully imposed under DBA. Therefore, in light of the above reasoning, the effectiveness of the proposed method, which relies on inducing an increase in double- and error-click rates, remains essentially unaffected by the intrinsic Geiger-mode deviations.

Thus, we consider the presented results to provide both validation of the countermeasure’s effectiveness and a proof-of-principle demonstration of the underlying theoretical framework.

Regarding the theoretical framework, it is useful to consider a possible implementation of the countermeasure where the detection probability depends on a randomly chosen supply voltage instead of switching between two fixed voltage levels. At first glance, this approach might seem to further limit the eavesdropper’s knowledge, as selecting a trigger pulse energy that remains undetected would become even less predictable.

However, this modification does not provide a fundamental security advantage. If the gate amplitude is reduced further, traces of Eve’s presence will still arise only when excessive optical power is applied during operation at the default voltage level. The double-click events will persist, while the secret key rate will inevitably decrease due to the reduced detection efficiency.

Conversely, if the random selection of the gate amplitude occasionally results in values sufficiently close to the default-voltage setting, condition~(\ref{eq:refname1}) may no longer be strictly satisfied. This reduces the fraction of bits identified as potentially compromised, leaving part of the privacy-amplified key available to Eve.

Finally, implementing a fully randomized gate-amplitude scheme would substantially complicate the practical realization of the countermeasure without yielding a proportional gain in security. For these reasons, the use of two fixed voltage levels represents a sufficient and practically optimal solution.

\section{Conclusions}

This study presents both a theoretical framework and empirical evidence demonstrating the efficacy of the proposed countermeasure against DBA. The method leverages statistical analysis of accumulated double-click and error events observed during a QKD session under randomized modulation of SPADs' detection probabilities via gate voltage manipulation. 

We tested random bias- and gate-voltage-level variations as countermeasures against click-imposition attacks under pulsed and CW DBA. While CW DBA results aligned with previous findings, new tests of countermeasures against pulsed DBA with a wide range of repetition rates have demonstrated a significant increase in the gap between $E^{Red. \, V}_{never}$ and $E^{Def. \, V}_{always}$. This widening of the energy gap correlates with a reduction in the average blinding power applied during the pulsed DBA. Notably, changing the gate voltage proved to be the only effective method for increasing $E_{always}$ by more than 13 dB while maintaining a non-zero detection probability for unblinded detectors.
 
The required condition $E_{always}^{Def. \, V} \leq \frac{1}{2}E_{never}^{Red. \, V}$ for an eavesdropper leaving "fingerprints" (double- and error-clicks) is satisfied by a large margin. This allowed us to assess the fraction of bits affected by Eve in both 2- and 1-SPAD-based QKD systems. Our approach firstly limits the amount of information Eve can obtain while being undetected, and secondly allows secret key distribution under pulsed DBA.

Furthermore, we evaluated the applicability of the proposed countermeasure against pulsed DBA for real-world QKD systems. This involved running numerical simulations of the secure key fraction and SKR for a QKD system with random variation of detection probabilities and comparing the results with the standard decoy-state BB84. The simulations were conducted across various detection probabilities, similar to those investigated in our gap-increase phenomenon studies, while also accounting for optical system imperfections inherent to practical QKD systems. The results show that, for a 100 km 2-SPAD-based QKD system, Eve cannot compromise more than roughly 1/3 of the distributed key without being detected. But using the mathematical apparatus presented in this paper, legitimate users can carefully assess this fraction and eliminate it by applying a privacy-amplification procedure. Thus, the countermeasure provides secure QKD while minimally impacting the secret key generation rate. However, 1-SPAD-based QKD systems require lower dark count rates and superior optical alignment to achieve comparable secure key generation rates.

We further verified that our assessments remain conservative against a soft-attack strategy, in which Eve lowers the trigger pulse energy below $2E_{always}^{Def.\,V}$ to suppress double-click events at the expense of partial clicks imposition. We showed analytically that, throughout the admissible range of the default gate voltage probability $\alpha$ (including $\alpha_{opt}$), the imposed-bit fraction estimated under such an attack never exceeds the one obtained for the full-energy strategy considered in Section \ref{fraction estimation}, for both 2- and 1-SPAD-based systems. Consequently, the proposed estimate provides an upper bound on Eve's information regardless of the trigger pulse energy she selects.

Nevertheless, the CW detector blinding attack almost negates the effect of this defense, meaning the system still needs some basic detector current control. Therefore, the aforementioned defense, when properly configured and combined with our countermeasure, can be considered a powerful tool for protecting QKD systems from any type of DBA.

Ultimately, we would like to emphasize that in this work, we focus on demonstrating a general method that can be applied to a broad class of commercially available detectors and QKD systems. This general applicability constitutes one of the key advantages of the proposed countermeasure against blinding attacks. Even if a designed state-of-art detector is vulnerable to fluctuations of parameters, the countermeasure could still be effective if the reduced gate voltage is chosen with an appropriate safety margin due to condition (\ref{eq:refname1}), ensuring proper operation of our method regardless of possible side effects, including potential memory effects.

While the development of a fully implemented SPAD with real-time random modulation of detection probability, or its integration into a complete QKD system, is beyond the scope of the present studies, we hope that the proposed approach will stimulate further development and practical implementation by detector manufacturers and system designers.

\section*{Acknowledgements} \label{sec:acknowledgements}
The authors gratefully acknowledge the fruitful discussions with S.N. Molotkov and R.Y. Lokhmatov.

\appendix*\label{Appendix}
\appendix

\section{The derivation of double and successfully-imposed clicks probability} \label{A}

Here, we provide explicit derivation of the (\ref{eq:refname5}) and (\ref{eq:refname6}). Taking into account the experimental results obtained in Sec.\ref{results}, the conditional detection probabilities of causing Bob's detector click are as follows: 
\begin{align}
P_{\eta_1}^{f}|P_{high} = P_{\eta_2}^{f}|P_{high} = & P_{\eta_1}^{h}|P_{high} = P_{\eta_1}^{f}|P_{low} = 1,\nonumber\\ P_{\eta_2}^{h}|P_{high} = P_{\eta_2}^{f}|P_{low} = & P_{\eta_1}^{h}|P_{low} = 
P_{\eta_2}^{h}|P_{low} = 0.
\label{eq:refnamea1}
\end{align}
In (\ref{eq:refnamea1}), the following designations were used:

$P_{\eta_1}^{f}|P_{high}$ ($P_{\eta_2}^{f}|P_{high}$) - the conditional detection probability given a high-power trigger pulse applied during a default (lowered) gate, under the condition that Eve's and Bob's bases are matched;

$P_{\eta_1}^{h}|P_{high}$ ($P_{\eta_2}^{h}|P_{high}$) - the conditional detection probability given a high-power trigger pulse applied during a default (lowered) gate, under the condition that Eve's and Bob's bases are mismatched;

$P_{\eta_1}^{f}|P_{low}$ ($P_{\eta_2}^{f}|P_{low}$) - the conditional detection probability given a low-power trigger pulse applied during a default (lowered) gate, under the condition that Eve's and Bob's bases are matched;

$P_{\eta_1}^{h}|P_{low}$ ($P_{\eta_2}^{h}|P_{low}$) - the conditional detection probability given a low-power trigger pulse applied during a default (lowered) gate, under the condition that Eve's and Bob's bases are mismatched.

Hence, the probability of causing a double-click event on each pulse sent by Eve is:

\begin{align}
P_{double}^{bld} = &\frac{\alpha}{2} \biggl( P_{high} \times {P_{\eta_1}^{h}|P_{high}}^2 + P_{low} \times {P_{\eta_1}^{h}|P_{low}}^2\biggr) + \nonumber\\
&\frac{1 - \alpha}{2} \biggl(P_{high} \times {P_{\eta_2}^{h}|P_{high}}^2 + P_{low} \times {P_{\eta_2}^{h}|P_{low}}^2\biggr).
\label{eq:refnamea2}
\end{align}

Therefore, taking into account (\ref{eq:refnamea1}), (\ref{eq:refnamea2}), $P_{high} = \beta$ and $P_{low} = 1- \beta$ the double clicks' imposition probability to the total number of fake-states sent by Eve $N_{sent}$ can be assessed:
\begin{equation}
P_{double}^{bld}= \frac{\alpha \beta}{2}.
\label{eq:refnamea3}   
\end{equation}

Assuming the detectors are identical, the probability of successful imposition ($P_{2}^{suc}$) can be represented  similarly to (\ref{eq:refnamea3}) by the following expression:
\begin{align}
P_{2}^{suc} &=\frac{1}{2} \bigg( \alpha P_{high} (P_{\eta_1}^{f}|P_{high} + P_{\eta_1}^{h}|P_{high} - {P_{\eta_1}^{h}|P_{high}}^{2}) +\nonumber\\
&\alpha P_{low} (P_{\eta_1}^{f}|P_{low} +P_{\eta_1}^{h}|P_{low} - {P_{\eta_1}^{h}|P_{low}}^{2}) +\nonumber \\
&(1- \alpha) P_{high} (P_{\eta_2}^{f}|P_{high} + P_{\eta_2}^{h}|P_{high} - {P_{\eta_2}^{h}|P_{high}}^{2}) +\nonumber \\
&(1-\alpha) P_{low} (P_{\eta_2}^{f}|P_{low} +P_{\eta_2}^{h}|P_{low} - {P_{\eta_2}^{h}|P_{low}}^{2})\bigg) =\nonumber\\
&=\frac{\alpha + \beta - \alpha \beta}{2}.
\label{eq:refnamea4}
\end{align} 

\section{Estimation of the secure bits fraction for one-SPAD-based QKD} \label{B}

\begin{figure*}[ht]
  \begin{subfigure}[b]{0.48\textwidth}
    \includegraphics[width=\textwidth]{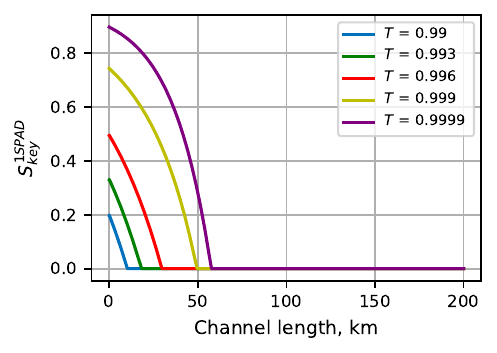}
  \end{subfigure}
  \begin{subfigure}[b]{0.49\textwidth}
    \includegraphics[width=\textwidth]{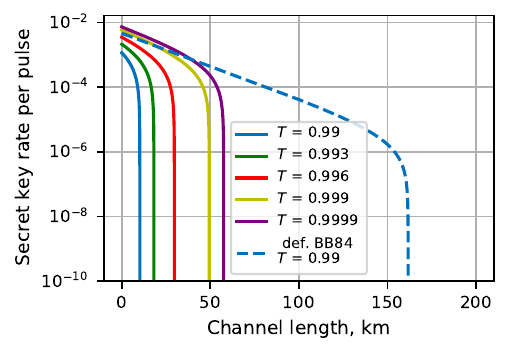}
  \end{subfigure}
  \caption{Secure bit fraction (left figure) and SKR per pulse (right figure) versus QKD channel length for a 1-SPAD QKD system under DBA assuming Eve mimics total and error click gains. Solid curves correspond to the proposed countermeasure for different extinction ratios (T). The dotted line shows the BB84 SKR for T = 0.99 with the detector gate voltage kept unchanged ($\eta = \eta_{1}$)}
\label{fig:7}
\end{figure*}

Likewise to the 2-SPAD-based approach, we first determine the probability of causing an error under pulsed DBA for a 1-SPAD-based QKD system ($P_{1}^{err}$) based on the estimated number of errors after the bases reconciliation procedure ($N^{exp}_{error}$):

\begin{align}
P_{1}^{err} =&\frac{1}{4} \biggl( \alpha P_{high} P_{\eta_1}^{h}|P_{high}  + (1 - \alpha) P_{high} P_{\eta_2}^{h}|P_{high}
+\nonumber\\
&\alpha P_{low} P_{\eta_1}^{h}|P_{low}+ (1 - \alpha) P_{low} P_{\eta_2}^{h}|P_{low} \biggr)=
\nonumber\\
=&\frac{\alpha \beta}{4} = \frac{1}{2}P_{double}^{bld} = \frac{2N^{exp}_{error}}{N_{sent}}.
\label{eq:refnameb1}
\end{align}

Secondly, in the case of a QKD system employing a single detector, the probability of successful click imposition can be expressed as:

\begin{align}
P_{1SPAD}^{succ} = &\frac{1}{4} \bigg( \alpha P_{high} \times (P_{\eta_1}^{f}|P_{high} + P_{\eta_1}^{h}|P_{high}) 
+\nonumber\\
&\alpha P_{low} \times (P_{\eta_1}^{f}|P_{low} +P_{\eta_1}^{h}|P_{low}) +\nonumber \\
&(1- \alpha) P_{high} \times (P_{\eta_2}^{f}|P_{high} + P_{\eta_2}^{h}|P_{high})
+\nonumber\\
&(1-\alpha) P_{low} \times (P_{\eta_2}^{f}|P_{low} +P_{\eta_2}^{h}|P_{low}) \bigg) =\nonumber\\
=&\frac {\alpha + \beta}{4} =\frac{N_{success}^{one}}{N_{sent}}.
\label{eq:refnameb2}
\end{align}

From (\ref{eq:refnameb1}) we can also estimate the fraction of trigger pulses Eve sends to induce clicks under lower supply voltage, corresponding to $N^{exp}_{error}$:

\begin{align}
\beta = \frac{8N^{exp}_{error}}{\alpha N_{sent}}.
\label{eq:refnameb3}
\end{align}

The number of successfully imposed bits $N_{success}^{one}$ obtained from (\ref{eq:refnameb2}) and (\ref{eq:refnameb3}) is:

\begin{align}
N_{success}^{one} = \frac{\alpha N_{sent}}{4} + \frac{2N^{exp}_{error}}{\alpha}.
\label{eq:refnameb4}
\end{align}

Hence, the total fraction of clicks unexposed to Eve ($S_{key}^{1SPAD}$) can be evaluated as:

\begin{align}
S_{key}^{1SPAD} &= \frac{N_{clicked}^{exp} - N_{success}^{one} -2N^{exp}_{error}}{N_{clicked}^{exp}} =\nonumber \\
&= 1 - \frac{\alpha N_{sent}}{4N_{clicked}^{exp}} - \frac{2N^{exp}_{error}(1+\alpha)}{\alpha N_{clicked}^{exp}}.
\label{eq:refnameb5}
\end{align}

For numerical simulations of the unattacked key fraction in a 1-SPAD receiver QKD system, we substitute $Q^{pass}$ with $\nicefrac{1}{2}Q^{pass}$ in (\ref{eq:refname14}), reflecting that Bob successfully registers only half the bits. To highlight the negative influence of receiver optical misalignment, we consider higher extinction ratios than for 2-SPAD-based QKD systems. We also assume Eve introduces a typical quantum bit error rate (QBER).

QBER is determined by optical system misalignment and dark count rate set by the following expression:

\begin{align}
    QBER =& \frac{1}{2} \biggl( \alpha \sum \limits _{i=0}^{2} n_{i} Q^{\mu_{i}(1-T)}_{\eta_{1}} + \nonumber \\
    & (1-\alpha)\sum \limits _{i=0}^{2} n_{i} Q^{\mu_{i}(1-T)}_{\eta_{2}}\biggr).
    \label{eq:refnameb6}
\end{align}

Thus, 
\begin{align}
N_{error}^{exp} = QBER * N_{Alice}.
\label{eq:refnameb7}
\end{align}

Following a similar approach outlined in Sec.\ref{fraction estimation}, we treat $N_{error}^{exp}$ as a "black-box" parameter. Consequently, the optimal $\alpha$ value for minimizing the estimated number of bits accessible to Eve when conducting DBA is:
\begin{align}
\alpha = 2\sqrt{\frac{2N_{error}^{exp}}{N_{Alice}Q^{Eve}}}.
\label{eq:refnameb8}
\end{align}

Therefore, $S_{key}^{1SPAD}$ corresponding to $\alpha$ from (\ref{eq:refnameb8}) is represented by the following expression:
\begin{align}
S_{key}^{1SPAD} = 1 - \frac{\sqrt{2N_{error}^{exp}N_{Alice}Q^{Eve}}}{N_{clicked}^{exp}} - \frac{2N_{error}^{exp}}{N_{clicked}^{exp}}.
\label{eq:refnameb9}
\end{align}

The left panel of Figure~\ref{fig:7} presents the simulated fraction of unaffected key bits, $S_{key}^{1SPAD}$) for $\eta_{2} = 0.108$ and various optical system extinction ratios (T).

Similary to (\ref{eq:refname17}), the lower bound on the SKR for the 1-SPAD-based QKD system is given by:

\begin{equation}
    \begin{aligned}
        R^{L}_{DBA} = &\frac{1}{4} S_{key}^{1SPAD}\biggl( \mu_{0} e^{-\mu_{0}} Y_{1}^{L}[1-H_{2}(e_{1}^{U})] \\
        &-Q_{\eta_{1},\eta_{2}}^{\mu_{0}} f_{EC} H_2(E_{\eta_{1},\eta_{2}}^{\mu_{0}}) \biggr).
        \label{eq:refnameb10}
    \end{aligned}
\end{equation}
The definitions and numerical values of the parameters are the same as in (\ref{eq:refname17}). The estimated SKR for $\eta_{2} = 0.108$ and various T, together with its comparison to the standard BB84 protocol (without the proposed countermeasure for $\eta_{1} =\eta_{2} = 0.121$), is presented in the right panel of Figure \ref{fig:7}.

The comparison of the estimated number of legitimate bits for 1- and 2-SPAD-based QKD systems reveals that imperfections in the receiver's optical and detection systems have a significantly more pronounced impact on 1-SPAD-based systems than on their 2-SPAD-based counterparts. This stems from the much lower intrinsic rate of double-click events compared to errors. Thus, being unable to distinguish error clicks caused by Eve from those arising due to system misalignment and dark counts, we regard all these events as the violator's influence. Therefore, 1-SPAD systems require a reduction in dark count rate and improvement of extinction ratio by several orders of magnitude to achieve comparable performance.

Nevertheless, similar to the 2-SPAD-based QKD systems, the number of error clicks follows a binomial distribution, approximating a normal distribution for large numbers. Given, that $\beta$-value $\sim$ $\nicefrac{N_{error}^{exp}}{N_{sent}}$ (\ref{eq:refnameb3}), the standard deviation of $\beta$ $\sigma_{\beta} \sim \nicefrac{1}{\sqrt{N_{sent}}}$. Therefore, considering statistical fluctuations will have an insignificant impact on the fraction of legitimate bits.

Now we apply the same reasoning to the soft-attack scenario considered in Section~\ref{soft-attack}. In the first soft-attack imposition energies subrange, when Eve's and Bob's bases are mismatched, the energy of the split pulse remains sufficient to trigger a default-gate detector with probability $\gamma$. Therefore, ($P_{\eta_1}^{h}|P_{high}=\gamma$) instead of unity in (\ref{eq:refnameb1}). Consequently, the error probability and the successful-imposition probability for a 1-SPAD-based system become

\begin{align}
P_{1,soft}^{err} &= \frac{\alpha \gamma}{4} 
= \frac{2N^{exp}_{error}}{N_{sent}},
\label{eq:refnameb11}\\
P_{1SPAD,soft}^{succ} &= \frac{\alpha(1 + \gamma)}{4}
= \frac{N_{success}^{one}}{N_{sent}}.
\label{eq:refnameb12}
\end{align}

Analogously to the 2-SPAD-based case, the number of successfully imposed clicks is given by:
\begin{align}
    N_{success,soft}^{one} = \frac{\alpha N_{sent}}{4} +2 N_{error}^{exp}.
    \label{eq:refnameb13}  
\end{align}

Comparing (\ref{eq:refnameb4}) and (\ref{eq:refnameb13}) shows that the estimate given by (\ref{eq:refnameb5}) remains the more conservative one for any value of $\alpha$.

In the second subrange, the split pulse energy is insufficient to generate a click, i.e., $\gamma = 0$. Consequently, the successful-imposition probability in the soft-attack scenario can never exceed the one obtained for the case considered first. 

Finally, in the full soft-attack scenario, Eve generates no clicks during the reduced-gate intervals. The resulting decrease in the gain associated with the lower supply voltage may therefore reveal her presence. A partial attack, in which only a fraction of pulses is intercepted, further reduces the number of Eve-induced clicks. Therefore, analogously to the 2-SPAD-based case, all soft-attack variants are fully encompassed by the estimate derived for the 1-SPAD-based QKD system.

\bibliography{Countermeasure.bib}

@article{wootters1982single,
  title={A single quantum cannot be cloned},
  author={Wootters, William K and Zurek, Wojciech H},
  journal={Nature},
  volume={299},
  number={5886},
  pages={802--803},
  year={1982},
  publisher={Nature Publishing Group UK London}
}

@article{gisin2006trojan,
  title={Trojan-horse attacks on quantum-key-distribution systems},
  author={Gisin, Nicolas and Fasel, Sylvain and Kraus, Barbara and Zbinden, Hugo and Ribordy, Gr{\'e}goire},
  journal={Physical Review A—Atomic, Molecular, and Optical Physics},
  volume={73},
  number={2},
  pages={022320},
  year={2006},
  publisher={APS}
}

@article{bugge2014laser,
  title={Laser damage helps the eavesdropper in quantum cryptography},
  author={Bugge, Audun Nystad and Sauge, Sebastien and Ghazali, Aina Mardhiyah M and Skaar, Johannes and Lydersen, Lars and Makarov, Vadim},
  journal={Physical review letters},
  volume={112},
  number={7},
  pages={070503},
  year={2014},
  publisher={APS}
}

@article{alferov2022study,
  title={Study of the vulnerability of neutral optical filters used in quantum key distribution systems against laser damage attack},
  author={Alferov, Sergei Vladimirovich and Bugai, Kirill Evgen'evich and Pargachev, Ivan Andreevich},
  journal={JETP Letters},
  volume={116},
  number={2},
  pages={123--127},
  year={2022},
  publisher={Springer}
}

@article{lydersen2010hacking,
  title={Hacking commercial quantum cryptography systems by tailored bright illumination},
  author={Lydersen, Lars and Wiechers, Carlos and Wittmann, Christoffer and Elser, Dominique and Skaar, Johannes and Makarov, Vadim},
  journal={Nature photonics},
  volume={4},
  number={10},
  pages={686--689},
  year={2010},
  publisher={Nature Publishing Group}
}

@article{sauge2011controlling,
  title={Controlling an actively-quenched single photon detector with bright light},
  author={Sauge, Sebastien and Lydersen, Lars and Anisimov, Andrey and Skaar, Johannes and Makarov, Vadim},
  journal={Optics Express},
  volume={19},
  number={23},
  pages={23590--23600},
  year={2011},
  publisher={Optical Society of America}
}

@article{huang2019laser,
  title={Laser-seeding attack in quantum key distribution},
  author={Huang, Anqi and Navarrete, {\'A}lvaro and Sun, Shi-Hai and Chaiwongkhot, Poompong and Curty, Marcos and Makarov, Vadim},
  journal={Physical Review Applied},
  volume={12},
  number={6},
  pages={064043},
  year={2019},
  publisher={APS}
}

@article{wiechers2011after,
  title={After-gate attack on a quantum cryptosystem},
  author={Wiechers, Carlos and Lydersen, Lars and Wittmann, Christoffer and Elser, Dominique and Skaar, Johannes and Marquardt, Ch and Makarov, Vadim and Leuchs, Gerd},
  journal={New Journal of Physics},
  volume={13},
  number={1},
  pages={013043},
  year={2011},
  publisher={IOP Publishing}
}

@article{makarov2006effects,
  title={Effects of detector efficiency mismatch on security of quantum cryptosystems},
  author={Makarov, Vadim and Anisimov, Andrey and Skaar, Johannes},
  journal={Physical Review A—Atomic, Molecular, and Optical Physics},
  volume={74},
  number={2},
  pages={022313},
  year={2006},
  publisher={APS}
}

@article{qi2005time,
  title={Time-shift attack in practical quantum cryptosystems},
  author={Qi, Bing and Fung, Chi-Hang Fred and Lo, Hoi-Kwong and Ma, Xiongfeng},
  journal={arXiv preprint quant-ph/0512080},
  year={2005}
}

@article{lydersen2011controlling,
  title={Controlling a superconducting nanowire single-photon detector using tailored bright illumination},
  author={Lydersen, Lars and Akhlaghi, Mohsen K and Majedi, A Hamed and Skaar, Johannes and Makarov, Vadim},
  journal={New Journal of Physics},
  volume={13},
  number={11},
  pages={113042},
  year={2011},
  publisher={IOP Publishing}
}

@article{jouguet2013preventing,
  title={Preventing calibration attacks on the local oscillator in continuous-variable quantum key distribution},
  author={Jouguet, Paul and Kunz-Jacques, S{\'e}bastien and Diamanti, Eleni},
  journal={Physical Review A—Atomic, Molecular, and Optical Physics},
  volume={87},
  number={6},
  pages={062313},
  year={2013},
  publisher={APS}
}

@inproceedings{bogdanov2022influence,
  title={Influence of QKD apparatus parameters on the backflash attack},
  author={Bogdanov, Sergey A and Sushchev, Ivan S and Klimov, Andrey N and Bugai, Klimov E and Bulavkin, DS and Dvoretsky, DA},
  booktitle={Quantum Technologies 2022},
  volume={12133},
  pages={90--95},
  year={2022},
  organization={SPIE}
}

@article{sushchev2024trojan,
  title={Trojan-horse attack on a real-world quantum key distribution system: Theoretical and experimental security analysis},
  author={Sushchev, Ivan S and Bulavkin, Daniil S and Bugai, Kirill E and Sidelnikova, Anna S and Dvoretskiy, Dmitriy A},
  journal={Physical Review Applied},
  volume={22},
  number={3},
  pages={034032},
  year={2024},
  publisher={APS}
}

@inproceedings{bugai2024protection,
  title={Protection Method against Powerful Emission Attacks Based on Optical-Fiber Fuse Element},
  author={Bugai, KE and Sushchev, IS and Bulavkin, DS and Lokhmatov, R Yu and Dvoretskiy, DA},
  booktitle={2024 International Conference Laser Optics (ICLO)},
  pages={446--446},
  year={2024},
  organization={IEEE}
}

@article{lovic2023quantified,
  title={Quantified effects of the laser-seeding attack in quantum key distribution},
  author={Lovic, Victor and Marangon, Davide G and Smith, PR and Woodward, Robert I and Shields, Andrew J},
  journal={Physical Review Applied},
  volume={20},
  number={4},
  pages={044005},
  year={2023},
  publisher={APS}
}

@article{lydersen2010thermal,
  title={Thermal blinding of gated detectors in quantum cryptography},
  author={Lydersen, Lars and Wiechers, Carlos and Wittmann, Christoffer and Elser, Dominique and Skaar, Johannes and Makarov, Vadim},
  journal={Optics express},
  volume={18},
  number={26},
  pages={27938--27954},
  year={2010},
  publisher={Optical Society of America}
}

@article{makarov2009controlling,
  title={Controlling passively quenched single photon detectors by bright light},
  author={Makarov, Vadim},
  journal={New Journal of Physics},
  volume={11},
  number={6},
  pages={065003},
  year={2009},
  publisher={IOP Publishing}
}

@inproceedings{bulavkin2023study,
  title={Study of a single-photon detector blinding attack with modulated bright light},
  author={Bulavkin, DS and Sushchev, IS and Bugai, KE and Bogdanov, SA and Dvoretskiy, DA},
  booktitle={Quantum and Nonlinear Optics IX},
  volume={12323},
  pages={73--78},
  year={2023},
  organization={SPIE}
}

@article{wu2020hacking,
  title={Hacking single-photon avalanche detectors in quantum key distribution via pulse illumination},
  author={Wu, Zhihao and Huang, Anqi and Chen, Huan and Sun, Shi-Hai and Ding, Jiangfang and Qiang, Xiaogang and Fu, Xiang and Xu, Ping and Wu, Junjie},
  journal={Optics Express},
  volume={28},
  number={17},
  pages={25574--25590},
  year={2020},
  publisher={Optical Society of America}
}

@article{yuan2011resilience,
  title={Resilience of gated avalanche photodiodes against bright illumination attacks in quantum cryptography},
  author={Yuan, ZL and Dynes, JF and Shields, AJ},
  journal={Applied physics letters},
  volume={98},
  number={23},
  year={2011},
  publisher={AIP Publishing}
}

@article{chistiakov2019controlling,
  title={Controlling single-photon detector ID210 with bright light},
  author={Chistiakov, Vladimir and Huang, Anqi and Egorov, Vladimir and Makarov, Vadim},
  journal={Optics express},
  volume={27},
  number={22},
  pages={32253--32262},
  year={2019},
  publisher={Optical Society of America}
}

@article{qian2019robust,
  title={Robust countermeasure against detector control attack in a practical quantum key distribution system},
  author={Qian, Yong-Jun and He, De-Yong and Wang, Shuang and Chen, Wei and Yin, Zhen-Qiang and Guo, Guang-Can and Han, Zheng-Fu},
  journal={Optica},
  volume={6},
  number={9},
  pages={1178--1184},
  year={2019},
  publisher={Optical Society of America}
}

@article{wu2020robust,
  title={Robust countermeasure against detector control attack in a practical quantum key distribution system: comment},
  author={Wu, Zhihao and Huang, Anqi and Qiang, Xiaogang and Ding, Jiangfang and Xu, Ping and Fu, Xiang and Wu, Junjie},
  journal={Optica},
  volume={7},
  number={10},
  pages={1391--1393},
  year={2020},
  publisher={Optical Society of America}
}

@article{acheva2023automated,
  title={Automated verification of countermeasure against detector-control attack in quantum key distribution},
  author={Acheva, Polina and Zaitsev, Konstantin and Zavodilenko, Vladimir and Losev, Anton and Huang, Anqi and Makarov, Vadim},
  journal={EPJ Quantum Technology},
  volume={10},
  number={1},
  pages={22},
  year={2023},
  publisher={Springer Berlin Heidelberg}
}

@article{gras2021countermeasure,
  title={Countermeasure against quantum hacking using detection statistics},
  author={Gras, Ga{\"e}tan and Rusca, Davide and Zbinden, Hugo and Bussi{\`e}res, F{\'e}lix},
  journal={Physical Review Applied},
  volume={15},
  number={3},
  pages={034052},
  year={2021},
  publisher={APS}
}

@article{shen2025countering,
  title={Countering detector manipulation attacks in quantum communication through detector self-testing},
  author={Shen, Lijiong and Kurtsiefer, Christian},
  journal={APL Photonics},
  volume={10},
  number={1},
  year={2025},
  publisher={AIP Publishing}
}

@article{lim2015random,
  title={Random variation of detector efficiency: A countermeasure against detector blinding attacks for quantum key distribution},
  author={Lim, Charles Ci Wen and Walenta, Nino and Legr{\'e}, Matthieu and Gisin, Nicolas and Zbinden, Hugo},
  journal={IEEE Journal of Selected Topics in Quantum Electronics},
  volume={21},
  number={3},
  pages={192--196},
  year={2015},
  publisher={IEEE}
}

@misc{legre2018apparatus,
  title={Apparatus and method for the detection of attacks taking control of the single photon detectors of a quantum cryptography apparatus by randomly changing their efficiency},
  author={Legre, Matthieu and Ribordy, Gr{\'e}gorie},
  year={2018},
  month=jul # "~10",
  publisher={Google Patents},
  note={US Patent 10,020,937}
}

@misc{bussieres2020blinding,
  title={Blinding attack detecting device and method},
  author={Bussi{\`e}res, F{\'e}lix and Ga{\"e}tan, GRAS},
  year={2020},
  month=aug # "~30",
  publisher={Google Patents},
  note={European Patent 3,716,252}
}

@article{kulik2017decoy,
  title={Decoy state method for quantum cryptography based on phase coding into faint laser pulses},
  author={Kulik, SP and Molotkov, SN},
  journal={Laser Physics Letters},
  volume={14},
  number={12},
  pages={125205},
  year={2017},
  publisher={IOP Publishing}
}

@misc{molotkov2021blinding,
  title={A method for detecting detector blinding attacks in quantum cryptography systems with polarization coding},
  author={Molotkov, Sergey},
  year={2021},
  month=nov # "~22",
  publisher={Google Patents},
  note={Russian Patent 2,783,977}
}

@article{huang2016testing,
  title={Testing random-detector-efficiency countermeasure in a commercial system reveals a breakable unrealistic assumption},
  author={Huang, Anqi and Sajeed, Shihan and Chaiwongkhot, Poompong and Soucarros, Mathilde and Legr{\'e}, Matthieu and Makarov, Vadim},
  journal={IEEE Journal of Quantum Electronics},
  volume={52},
  number={11},
  pages={1--11},
  year={2016},
  publisher={IEEE}
}

@article{ma2005practical,
  title={Practical decoy state for quantum key distribution},
  author={Ma, Xiongfeng and Qi, Bing and Zhao, Yi and Lo, Hoi-Kwong},
  journal={Physical Review A—Atomic, Molecular, and Optical Physics},
  volume={72},
  number={1},
  pages={012326},
  year={2005},
  publisher={APS}
}

@article{ma2002multiplication,
  title={Multiplication in separate absorption, grading, charge, and multiplication InP-InGaAs avalanche photodiodes},
  author={Ma, CLF and Deen, MJ and Tarof, LE},
  journal={IEEE Journal of Quantum Electronics},
  volume={31},
  number={11},
  pages={2078--2089},
  year={2002},
  publisher={IEEE}
}

@article{miller1955avalanche,
  title={Avalanche breakdown in germanium},
  author={Miller, SL},
  journal={Physical review},
  volume={99},
  number={4},
  pages={1234},
  year={1955},
  publisher={APS}
}

@article{cova1996avalanche,
  title={Avalanche photodiodes and quenching circuits for single-photon detection},
  author={Cova, Sergio and Ghioni, Massimo and Lacaita, Andrea and Samori, Carlo and Zappa, Franco},
  journal={Applied optics},
  volume={35},
  number={12},
  pages={1956--1976},
  year={1996},
  publisher={OSA}
}

@article{zhao2006experimental,
  title={Experimental quantum key distribution with decoy states},
  author={Zhao, Yi and Qi, Bing and Ma, Xiongfeng and Lo, Hoi-Kwong and Qian, Li},
  journal={Physical review letters},
  volume={96},
  number={7},
  pages={070502},
  year={2006},
  publisher={APS}
}

@article{kravtsov2018relativistic,
  title={Relativistic quantum key distribution system with one-way quantum communication},
  author={Kravtsov, KS and Radchenko, IV and Kulik, SP and Molotkov, SN},
  journal={Scientific reports},
  volume={8},
  number={1},
  pages={6102},
  year={2018},
  publisher={Nature Publishing Group UK London}
}

@article{restelli2012time,
  title={Time-domain measurements of afterpulsing in InGaAs/InP SPAD gated with sub-nanosecond pulses},
  author={Restelli, Alessandro and Bienfang, Joshua C and Migdall, Alan L},
  journal={Journal of Modern Optics},
  volume={59},
  number={17},
  pages={1465--1471},
  year={2012},
  publisher={Taylor \& Francis}
}

@article{restelli2013single,
  title={Single-photon detection efficiency up to 50\% at 1310 nm with an InGaAs/InP avalanche diode gated at 1.25 GHz},
  author={Restelli, Alessandro and Bienfang, Joshua C and Migdall, Alan L},
  journal={Applied Physics Letters},
  volume={102},
  number={14},
  year={2013},
  publisher={AIP Publishing}
}

@article{jain2017inp,
  title={InP/InAsP nanowire-based spatially separate absorption and multiplication avalanche photodetectors},
  author={Jain, Vishal and Heurlin, Magnus and Barrigon, Enrique and Bosco, Lorenzo and Nowzari, Ali and Shroff, Shishir and Boix, Virginia and Karimi, Mohammad and Jam, Reza J and Berg, Alexander and others},
  journal={ACS Photonics},
  volume={4},
  number={11},
  pages={2693--2698},
  year={2017},
  publisher={ACS Publications}
}

@article{williams2013multi,
  title={Multi-gain-stage InGaAs avalanche photodiode with enhanced gain and reduced excess noise},
  author={Williams, George M and Compton, Madison and Ramirez, David A and Hayat, Majeed M and Huntington, Andrew S},
  journal={IEEE Journal of the Electron Devices Society},
  volume={1},
  number={2},
  pages={54--65},
  year={2013},
  publisher={IEEE}
}

@article{losev2022dead,
  title={Dead time duration and active reset influence on the afterpulse probability of InGaAs/InP single-photon avalanche diodes},
  author={Losev, Anton V and Zavodilenko, Vladimir V and Koziy, Andrey A and Filyaev, Alexandr A and Khomyakova, Kristina I and Kurochkin, Yury V and Gorbatsevich, Alexander A},
  journal={IEEE Journal of Quantum Electronics},
  volume={58},
  number={3},
  pages={1--11},
  year={2022},
  publisher={IEEE}
}

@article{ma2005security,
  title={Security of quantum key distribution with realistic devices},
  author={Ma, Xiongfeng},
  journal={arXiv preprint quant-ph/0503057},
  year={2005}
}

\end{document}